\documentclass[aps,pre,twocolumn,groupedaddress]{revtex4-2}
\usepackage{amsmath,amssymb,mathtools}
\usepackage{slashed}
\usepackage{graphicx}
\graphicspath{ {./figures/} }
\usepackage{hyperref}
\usepackage{cleveref}
\usepackage{verbatim}
\usepackage{verbatim}

\usepackage{float}
\usepackage{subfig}
\usepackage{soul}

\usepackage[percent]{overpic}

\usepackage[dvipsnames]{xcolor}

\newcommand{\dd}{\text{d}}
\renewcommand{\vec}[1]{\boldsymbol{#1}} 

\begin{document}
\title{Snap, Crackle, and Pop: This is why the potential of mean force clashes with the fluctuation dissipation relation}

\author{Fabian Koch}
\email[]{fabian.glatzel@physik.uni-freiburg.de}
\author{Tabita Wasmer}
\author{Tanja Schilling}
\email[]{tanja.schilling@physik.uni-freiburg.de}
\affiliation{Institut f\"ur Physik, Albert-Ludwigs-Universit\"at Freiburg,\\ Hermann-Herder-Stra\ss e 3, 79104 Freiburg im Breisgau, Germany}

\date{\today}

\begin{abstract}
     We analyze the non-linear generalized Langevin equation which contains a thermodynamic force. We show that even for systems in thermal equilibrium the presence of the thermodynamic force implies that the auto-correlation function of the fluctuating force becomes non-stationary. We further illustrate that a standard coarse-graining procedure that neglects this fact predicts waiting-time distributions incompatible with the original, microscopic process. We conclude that one needs to proceed with care when adding thermodynamic driving forces to the Langevin equation.
\end{abstract} 
\maketitle

In physics, stochastic models are widely used. Whether we consider non-equilibrium work relations, the kinetics of phase transitions, or the properties of soft and biological materials, we encounter evolution equations of the Langevin type \cite{vankampen1992, chaikin1995, peters2017, snook2006, seifert2012, zwanzig2001}. In contrast, the equations of motion of microscopic, closed systems are deterministic. If a stochastic evolution equation is used to describe an open or coarse-grained system it needs to be compatible with the underlying deterministic microscopic dynamics. The term ``compatible'' here means that the stochastic equation needs to predict the same evolution as the microscopic equations of motion for averages of the coarse-grained variables and preferably also for certain correlation functions of these variables. 

While the heuristic justifications of stochastic evolution equations which were brought forward by Einstein, Langevin and their contemporaries appear very convincing, it is rather difficult to show rigorous proofs of the emergence of these equations from the Liouville-von Neumann equation \cite{Forster_1975_Hydrodynamic_book,Fick_1986_Quantenstatistik_book,Bocquet_1994_Brownian,givon2005,widder2025}.

In this letter, we focus on stochastic evolution equations which contain a thermodynamic driving force such as a derivative of a free energy. Such equations are often used when modeling the kinetics of phase transitions and of chemical reactions, e.g.~in the context of transition state theory, Markov state modeling and phase field modeling \cite{chaikin1995, peters2017}. Dynamic density functional theory, in which the variation of a free energy functional determines the stochastic dynamics of the density, is of a similar structure \cite{vrugt2020ddft}. Also in the context of materials simulations of soft matter the concept of a thermodynamic driving force is often used: The potential of mean force describes the effective interactions between coarse-grained units and, despite being of thermodynamic nature, is treated like a mechanical force in the stochastic evolution equations \cite{jin2022,Klimek_2025_Subdiffusion}.

Stochastic models are often motivated by their deterministic counterparts that can be derived from the underlying microscopic equations of motion. Consider as an example the evolution of one component (the $z$-component) of the momentum $\vec{p}$ of one particle out of a many particle system. In the literature on molecular dynamics \cite{Ruehle_2009_Versatile,Praprotnik_2008_Multiscale,Han_2018_Mesoscopic,Han_2021_Constructing,Groot_1997_Dissipative,Pivkin_2010_Dissipative,Espanol_2017_Perspective} one often finds such variables described by the non-linear generalized Langevin equation
\begin{equation} \label{eq:potMeanForce1}\frac{\dd p_z(t)}{\dd t} =  -\left.\frac{\dd W_\text{MF}(z)}{\dd z}\right|_{z=z(t)} - \int^t_0 K(t-s) p_z(s) \, \text{d}s \, + \eta(t) \, .
\end{equation}
Here, $z$ is the $z$-component of the position, $K(t)$ is a memory function, $\eta(t)$ is the "fluctuating force" and $W_\text{MF}(z)$ is the potential of mean force. $W_\text{MF}(z)$ is defined via the probability $P_{\rm eq}(z)$ of finding the particle at a position $z$ in the equilibrium ensemble, $W_\text{MF}(z) := -k_BT \ln P_{\rm eq}(z)$. (In other contexts, such as phase field modelling and transition state theory, this type of function is named "free energy landscape" $\Delta G(z)$.) 
 
 Additionally, the second fluctuation dissipation relation
(2FDT)
\begin{align}
    \label{eq:2FDT}
    K(t)&=
    \frac{\langle\eta(s+t)\,\eta(s)\rangle}{\langle p_z^2\rangle} \quad \forall s
\end{align}
is assumed to hold true \cite{Lei_2016_Data,Daldrop_2018_Butane,Kappler_2019_Cyclization,Wang_2019_Coarse} -- as it would for the linear version of the GLE. In order to obtain a stochastic evolution equation, the fluctuating force $\eta$ is then replaced by a suitable stochastic process $\xi$. However, this procedure relies on the validity of the 2FDT for non-linear versions of the GLE.  This point has been under debate recently \cite{glatzel2022interplay,vroylandt2022derivation,vroylandt2022position}.

The question which we would like to resolve in this letter is whether eqs.\ref{eq:potMeanForce1} and \ref{eq:2FDT} follow in general from the microscopic dynamics of an equilibrium system -- or, more specifically, whether it is correct to impose eq.~\ref{eq:2FDT} on the noise in eq.~\ref{eq:potMeanForce1}. Work from the early 2000s confirmed the compatibility of the potential of mean force both, with a linear memory term and with the 2FDT \cite{Lange2006,kinjo2007}. However, in both derivations incorrect transformations were applied to the projection operators. In 2014 Carof and co-workers remarked that the noise did not fulfill the 2FDT \cite{carof2014}. In 2022 we showed that the potential of mean force is neither compatible with a linear memory term nor with the 2FDT  \cite{glatzel2022interplay}, while at the same time Vroylandt and Monmarch\'e argued that the linear memory term was compatible with the potential of mean force, if one introduced an additional term to the 2FDT \cite{vroylandt2022position,vroylandt2022derivation}.

All derivations that have been brought forward so far in this debate suffer from one common shortcoming: they have been derived by means of the projection-operator formalism. This formalism has been well established and widely used since the 1960s \cite{Mori_Transport_1965, zwanzig1961, zwanzig2001, grabert2006projection, PMID:10737778}. It serves to systematically integrate out degrees of freedom of complex systems. Despite its obvious success, it has drawbacks: First, the formalism allows to construct exact evolution equations, but it does not allow to disprove equations that it cannot construct.
Second, its derivation contains a precarious step. The so-called ``orthogonal dynamics'', i.e.~the dynamics of those degrees of freedom that are integrated out, is handled by means of the Duhamel-Dyson identity. In physics textbooks (e.g.~ref.~\cite{zwanzig2001, snook2006, grabert2006projection}) it is not questioned that this identity can be used, although it is rather difficult to show under which conditions the orthogonal dynamics is a strongly continuous semigroup -- which is a prerequisite for the Duhamel-Dyson identity to hold \cite{givon2005,zhu2018,widder2025}. It has recently been proven  that the formalism is correct for projection operators of finite rank \cite{widder2025} (such as the "Mori projector" \cite{Mori_Transport_1965}). However, evolution equations with a thermodynamic driving force such as eq.~\ref{eq:potMeanForce1} require a projection operator of infinite rank (the "Zwanzig projector" \cite{zwanzig1961}). For this case to our knowledge there is no proof of the Duhamel-Dyson identity. Third, while there is no general derivation of the validity of the combination of eq.~\ref{eq:potMeanForce1} and eq.~\ref{eq:2FDT} by means of the projection operator formalism, there are specific model systems for which it can be proven (e.g.~the Caldeira-Leggett or Kac-Zwanzig model \cite{zwanzig2001}). These proofs seem to be mistaken by some researchers for proofs of the general case. 

To sidestep these problems, here we show an analysis of eqs.\ref{eq:potMeanForce1} and \ref{eq:2FDT}, which does not require the projection-operator formalism. We will assume that \cref{eq:potMeanForce1} holds and then test which implication this assumption has for \cref{eq:2FDT}.

First, we use \cref{eq:potMeanForce1,eq:2FDT} for $s=0$ to obtain the following Volterra equation for the memory kernel:
\begin{align}
    K(t)\langle p_z^2 \rangle &= \langle \dot{p}_z(t)\eta(0) \rangle - \langle F_\mathrm{MF}(z(t)) \xi(0) \rangle\nonumber\\
    &\phantom{=}+ \int_0^t \dd s\, K(t-s) \langle p_z(s) \xi(0) \rangle\, .\label{eq:kernel_volterra}
\end{align}
Here we have introduced the mean force $F_\text{MF}(z) \coloneqq -\dd W_\text{MF}(z)/\dd z$.

From \cref{eq:potMeanForce1} at $t=0$, we see that
\begin{align}
    \eta(0) &= \chi(0)
\end{align}
with $\chi(t):=\dot{p}_z(t)-F_\text{MF}(z(t))$. All quantities in \cref{eq:kernel_volterra} except for $K(t)$ can directly be obtained from measurements or simulation data. If the correlation functions appearing in \cref{eq:kernel_volterra} are continuous, then due to the existence and uniqueness of the solutions of Volterra equations, \cref{eq:kernel_volterra} can be used as a definition for $K(t)$. With $K(t)$ at hand, \cref{eq:potMeanForce1} uniquely defines $\eta(t)$.
As \cref{eq:2FDT} for $s=0$ is used to define $K(t)$, $\eta(t)$ fulfills this relation by construction for $s=0$. Now we will check if \cref{eq:2FDT} also holds for $s\neq 0$. This is the case if and only if
\begin{align}
    \frac{\dd}{\dd s} \langle\eta(s+t)\eta(s)\rangle &\overset{!}{=} 0\quad\forall s,t.\label{eq:stationarity_condition}
\end{align}
To phrase this condition in different terms, we first consider the action of the time-evolution operator $\mathcal{U}(s)$ on $\eta(t)$. As $\eta(t)$ is not an observable the time evolution of which follows from the stream lines in phase space \cite{Mori_Transport_1965}, we use \cref{eq:potMeanForce1} to determine 
\begin{align}
    \mathcal{U}(s) \eta(t-s) &= \eta(t) - \int_0^{s} \, \dd s' K(t-s')p_z(s').
\end{align}
By differentiating this equation we find
\begin{align}
    \frac{\dd}{\dd s} \mathcal{U}(s)\eta(t-s) = -K(t-s)p_z(s).
\end{align}
This result can be used to rewrite \cref{eq:stationarity_condition} as
\begin{align}
    \frac{\dd}{\dd s} &\langle \eta(s+t)\eta(s)\rangle\! = \frac{\dd}{\dd s} \langle [\mathcal{U}(-s) \eta(t+s)] [\mathcal{U}(-s)\eta(s)]\rangle \\
    &= K(t+s)\langle p_z(0) \eta(s)\rangle + \langle\eta(t+s) p_z(0) \rangle K(s).
    \label{eq:timederivative}
\end{align}
Hence, \cref{eq:stationarity_condition} is fulfilled if $\langle\eta(t)p_z(0) \rangle=0\ \forall t$. By choosing $t=0$ in \cref{eq:timederivative}, we see that the implication also holds in the opposite direction as $K(s)\neq 0$. Hence, there is the equivalence
\begin{align}
\label{eq:equivalence}
    \frac{\dd}{\dd s}\langle\eta(s+t)\eta(s)\rangle &\equiv 0 &\Longleftrightarrow&&\langle\eta(t) p_z(0)\rangle\equiv0 \; .
\end{align}
In order to check whether $\langle\eta(t) p_z\rangle\equiv0$, we consider the Taylor series
\begin{align}
    \langle \eta(t)p_z(0)\rangle &=  \sum_{n=0}^3\frac{t^n}{n!}\left[\frac{\dd^n}{\dd t^n}\langle\eta(t)p_z(0)\rangle\right]_{t=0}+ \mathcal{O}(t^4).
\end{align}
If any of the Taylor coefficients is non-zero, the entire function cannot vanish for all $t$ and consequently \cref{eq:2FDT} cannot hold. In the following, we always assume that we deal with a canonical ensemble.

For the $n=0$ coefficent, one easily finds
\begin{align}
    \langle\eta(0)p_z(0)\rangle &= \langle \chi(0) p_z(0)\rangle=0
\end{align}
because $\chi$ does not depend on the microscopic momenta and, thus, the product $\chi(0)p_z(0)$ is a odd function of the momentum. By inserting \cref{eq:potMeanForce1} into the expression with $n=1$, we find
\begin{align}
    \frac{\dd}{\dd t}\langle\eta(t)p_z(0)\rangle &= \langle\eta(t)F_\text{MF}(z(0))\rangle.
\end{align}
The fluctuating force and the momentum are orthogonal if and only if the fluctuating force and the mean force are orthogonal. By using $\eta(0)=\chi(0)$ and expressing the ensemble averages in terms of conditional expectation values, one easily finds
\begin{align}
    \langle\eta(0)F_\text{MF}(z(0))\rangle &= 0
\end{align}
hence, also the second term in the Taylor expansion is zero. By a similar line of reasoning for $n=2$, we find that
\begin{align}
    \frac{\dd^2}{\dd t^2}\langle\eta(t)p_z(0)\rangle &= -\langle\eta(t)\mathcal{L}F_\text{MF}(z(0))\rangle
\end{align}
where $\mathcal{L}$ is the Liouvillian. At $t=0$ we obtain
\begin{align}
     -\langle\eta(0)\mathcal{L}F_\text{MF}(z(0))\rangle &= \langle(\mathcal{L}^2p_z)F_\text{MF}(z(0))\rangle=0
\end{align}
which is easily shown by rewriting the expression in terms of (conditional) expectation values. Hence, also the third term in the Taylor expansion vanishes. Finally, we regard $n=3$, i.e.~the fourth order term. Following the same procedure as before, we find
\begin{align}
    \frac{\dd^3}{\dd t^3}\langle\eta(t)p_z(0)\rangle &= \langle\eta(t)\mathcal{L}^2F_\text{MF}(z(0))\rangle+K(t)\langle F_\text{MF}(z(0))^2\rangle
\end{align}
and for $t=0$ we get
\begin{align}
    \frac{\dd^3}{\dd t^3}\langle\eta(t)p_z(0)\rangle\bigg|_{t=0} \!\!&= m\langle F_\text{MF}(z(0))\mathcal{L}^4z(0)\rangle\!\nonumber\\
    &\phantom{=}+\!\langle(\mathcal{L}F_\text{MF}(z(0)))^2\rangle\!+\!K(0)\langle F_\text{MF}(z)^2\rangle\label{eq:fourth_order_term}
\end{align}
which does not vanish in general. (The quantity $\mathcal{L}^4z$ rarely appears in physics. It is called called jounce or snap. The fifth and sixth order derivative of the position are sometimes humorously called crackle and pop \cite{Visser_2004,Eager_2016}. Thus we argue that snap, crackle and pop spoil the 2FDT.)

In summary, the right hand term in \cref{eq:equivalence} is not zero in general, which implies that \cref{eq:stationarity_condition} does not hold. This implies that \cref{eq:potMeanForce1} and \cref{eq:2FDT} are not compatible in general. I.e.~if one wishes to use a potential of mean force and a linear memory term in a Langevin model, then even in equilibrium the fluctuating forces are not stationary.  

As some readers might consider a non-stationary force-auto-correlation in an equilibrium system unphysical, we illustrate this statement by an example. In the following we will determine the fourth order term for a simple model system numerically.
Note that using \cref{eq:kernel_volterra} we can determine the value of $K(0)$ as
\begin{align}
    K(0) &= \frac{\langle\chi(0)^2\rangle}{\langle p_z(0)^2\rangle}
\end{align}
and, thus, all quantities in \cref{eq:fourth_order_term} are easily accessible in simulations. 

\begin{figure}[t]
    \centering
    \includegraphics[width= \columnwidth]{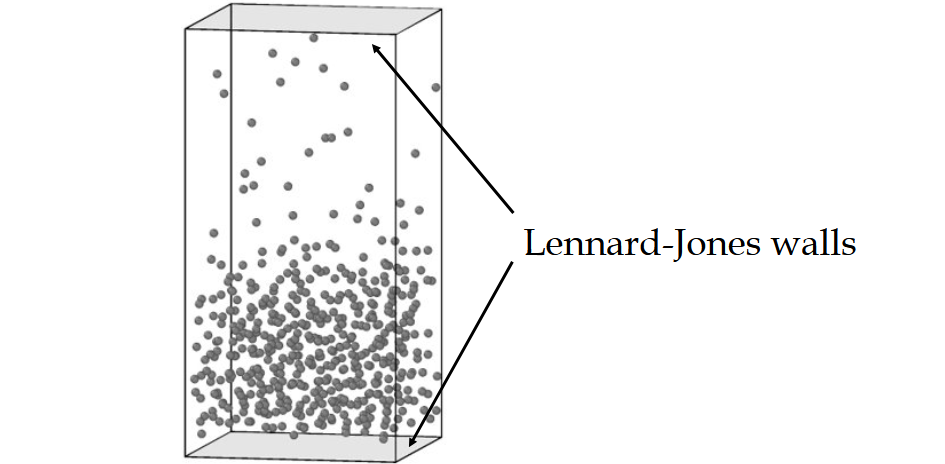}
    \caption{Snapshot of the simulated system with $500$ Lennard-Jones particles under gravity in a cuboid box}
    \label{fig:system_snapshot}
\end{figure}

 We analyze a system of $500$ identical Lennard-Jones (LJ) particles in a cuboid simulation box (see \cref{fig:system_snapshot}). Each particle has mass $m$. The LJ diameter and energy are denoted as $\sigma$ and $\epsilon$. The simulation box of the volume $V = 10\sigma \times 10\sigma \times 20 \sigma$ has periodic boundary conditions along the $x$ and $y$ direction and LJ-walls along the $z$ direction. Additionally, we have a constant force field along the $z$ direction acting on all the particles, i.e.~we simulate a closed ``ball pit'' under gravity. All interactions are modeled using truncated Lennard-Jones potentials with cutoff radius $r_\text{co} = 2.5 \sigma$.

 \begin{figure*}
  \includegraphics[width=\textwidth]{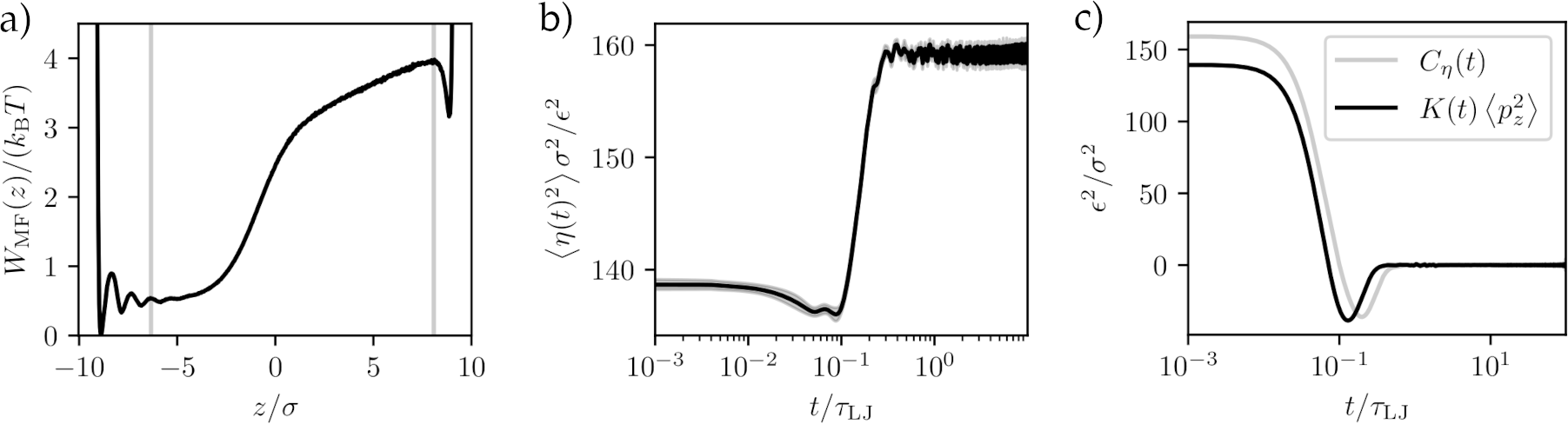}
  \caption{Results from the MD simulations of vertically confined Lennard-Jones particles under gravity. a) Potential of mean force with two vertical lines marking the region used for the analysis of the residence times. b) Variance of the fluctuating force as a function of time. The standard deviation of the mean is included as error band in light gray. c) Autocorrelation function of the fluctuating force in the stationary limit and the memory kernel (multiplied with $\langle p_z^2\rangle$).}
  \label{fig:md_results}
\end{figure*}

We perform molecular dynamics simulations with LAMMPS to propagate the system dynamics \cite{thompson2022lammps}. The equations of motion are integrated using the velocity Verlet algorithm with a time step of $\Delta t = 0.001 \tau_\text{LJ}$, where $\tau_\text{LJ} = \sqrt{m\sigma^2/\epsilon}$ is the characteristic Lennard-Jones time. To maintain a constant temperature $T$, we apply a Nosé-Hoover thermostat. We pick the $z$ component of the momentum of one particle as the observable of interest. By determining a histogram of the $z$ components of the positions of the particles after equilibration, we can measure $P_\text{eq}(z)$ and, thus, $W_\text{MF}(z)$ (see \cref{fig:md_results}a). We use umbrella sampling to improve the statistics at the boundaries. We then determine the mean force $F_\text{MF}(z)$ by numerically differentiating $W_\text{MF}(z)$ using a Savitzky-Golay filter to reduce noise while preserving relevant features. 

To evaluate the individual terms in \cref{eq:fourth_order_term}, we first express all quantities analytically by computing the action of the Liouvillian. In particular, for the first term related to the determination of snap, all microscopic forces acting on the particles are taken into account explicitly. While some momentum dependencies are already present in the original expressions, additional ones arise through the application of the Liouvillian. Subsequently, the ensemble average over momenta is carried out analytically under the assumption of a Maxwell-Boltzmann distribution for all momentum-dependent terms. This yields expressions, which are finally averaged over particle positions using simulation data. 
This procedure results in a numerical value for \cref{eq:fourth_order_term}, which is evaluated from the simulation data as
\begin{align}
    \frac{\dd^3}{\dd t^3}\langle\eta(t)p_z(0)\rangle\bigg|_{t=0} \!\!& = 2840 \pm 40 \, \frac{\epsilon^3}{\sigma^4 m}\neq 0. 
\end{align}

The non-stationary properties of the fluctuating force can also be seen more directly in the simulation data. To this end, we use \cref{eq:kernel_volterra} to determine the memory kernel from the data. Next, we use \cref{eq:potMeanForce1} to calculate $\eta(t)$ for the individual trajectories. With this, we can now determine $\langle\eta(0)\eta(t)\rangle$. As this $K(t)$ and $\langle\eta(0)\eta(t)\rangle$ do not satisfy the FKR/2FDT (cf. \cref{eq:2FDT}) perfectly, we make a new guess for a memory kernel by a linear combination of the original memory kernel and $\langle\eta(0)\eta(t)\rangle/\langle p_z^2\rangle$ and repeat the entire procedure until we obtain a self-consistent $K(t)$ and $\langle\eta(0)\eta(t)\rangle$ (see \cref{fig:md_results}c). With these values for $\eta(t)$, we determine $\langle \eta(t)^2\rangle$ as shown in \cref{fig:md_results}b. We can see that the variance of the fluctuating force depends explicitly on time and, thus, $\eta(t)$ cannot be modeled by a simple noise term with time-independent statistical properties. Further, we see that the variance of the fluctuating force has reached a plateau at around $t=1\tau_{\text{LJ}}$. We determine the autocorrelation function of $\eta(t)$ for times $t\geq 1\tau_{\rm LJ}$ and compare it with the memory kernel (see \cref{fig:md_results}c). We find that the FKR/2FDT is not fulfilled even at late times and that the mismatch cannot be resolved by a simple re-scaling (note that the positions of the minima of $C_\eta(t)$ and $K(t)$ do not match).

\begin{figure}
    \centering
    \includegraphics[scale=1]{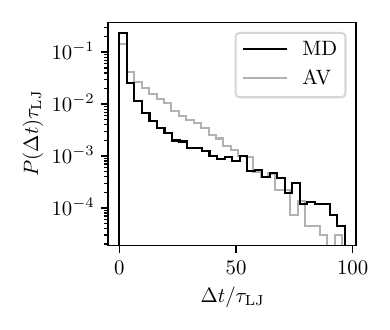}
    \caption{Distribution of residence times within the region in-between the two vertical lines in \cref{fig:md_results}a for the original molecular-dynamics (MD) simulations as well as the coarse-grained simulations via the auxiliary-variables (AV) method}
    \label{fig:residence_times}
\end{figure}

Finally, we illustrate that integrating the wrong coarse-grained equation of motion, i.e.~\cref{eq:potMeanForce1} with a noise term that satisfies the 2FDT at all times, does not reproduce some core properties of the original simulations. To this end, we integrate \cref{eq:potMeanForce1} with the auxiliary-variables method \cite{Shugard_1977_Dynamics,Doerries_2021_Correlation,Li_2017_Computing,Bruenig_2022_Time}. For the trajectories obtained this way as well as for the original MD trajectories we determine the residence-time distribution in the region between the two vertical lines in \cref{fig:md_results}a. \Cref{fig:residence_times} shows that the residence-time distributions deviate noticeably.

In this letter we have addressed the validity of the second fluctuation dissipation relation (2FDT) for the most common form of a non-linear generalized Langevin equation, the one that contains a derivative of a thermodynamic potential (i.e.~of a "free energy landscape") as a mean-force term. We have shown that a GLE of such a form, which fulfills the 2FDT initially, can be derived without the projection-operator formalism. However, the 2FDT is not fulfilled at later times in general. By a formal analysis of the autocorrelation function of the fluctuating force we show that there is no evidence that the 2FDT should be fulfilled at later times. We also confirmed this numerically. Hence, we conclude that for most systems a simple GLE with a mean-force term (i.e.~a derivative of a free energy) is incompatible with the standard 2FDT. Since considerable of effort is invested in optimizing algorithms to sample free energy landscapes and to propagate stochastic processes is these landscapes, we point out that often these equations will not describe the investigated process correctly. In particular, we showed that integrating a GLE with the 2FDT in a free energy landscape leads to wrong waiting time distributions even if the the landscape has been perfectly sampled.

\subsection*{Data Availability}
The data that support the findings of this study are available from the corresponding author upon reasonable request.

\subsection*{Acknowledgement}
The authors acknowledge funding by the Deutsche Forschungsgemeinschaft (DFG, German Research Foundation) in Project No. 431945604.

\bibliography{literature.bib}
\end{document}